\documentclass[secnumarabic,amssymb, amsmath, 
nofootinbib,tightenlines,
nobibnotes, aps, prl, preprint]{revtex4}
\usepackage{graphicx}
\catcode`@=11
\def\references{%
    \ifpreprintsty
    \bigskip\bigskip
    \hbox to\hsize{\hss\large \refname\hss}%
    \else
    \vskip 24pt
    \hrule width\hsize\relax
    \fi
    \list{\@biblabel{\arabic{enumiv}}}%
    {\labelwidth\WidestRefLabelThusFar  \labelsep4pt %
    \leftmargin\labelwidth %
    \advance\leftmargin\labelsep %
    \ifdim\baselinestretch pt>1 pt %
    \parsep  4pt\relax %
    \else %
    \parsep  0pt\relax %
    \fi
    \itemsep\parsep %
    \usecounter{enumiv}%
    \let\p@enumiv\@empty
    \def\theenumiv{\arabic{enumiv}}%
    }%
    \let\newblock\relax %
    \sloppy\clubpenalty4000\widowpenalty4000
    \sfcode`\.=1000\relax
    \ifpreprintsty\else\small\fi
}
\catcode`@=12

\def\gev{\rm GeV}
\def\fbi{\rm fb^{-1}}

\def\mumu{\mu^+\mu^-}

\def\lsim{\mathrel{\raise.3ex\hbox{$<$\kern-.75em\lower1ex\hbox{$\sim$}}}}
\def\gsim{\mathrel{\raise.3ex\hbox{$>$\kern-.75em\lower1ex\hbox{$\sim$}}}}

\begin{document}

\preprint{MAD-PH/01-1253}
\preprint{hep-ph/0201023}

\title {$h\to \mumu$ via gluon fusion at the LHC}
\author{Tao Han and Bob McElrath}
\affiliation{Department of Physics, University of Wisconsin--Madison, WI 53706}
\date{\today}

\begin{abstract} 
We study the observability of the $h\to \mumu$ decay in the Standard Model and
the MSSM at the LHC.  The observation of the $h\mu\mu$ coupling is important to
determine whether the Higgs particle that generates mass for the weak bosons is
also responsible for mass generation of the second generation of fermions.  We
find that the signal via the gluon fusion channel is comparable to that from
the weak-boson fusion.  By combining these two channels, observing $h\to \mumu$
is feasible at the LHC with a delivered luminosity of $300\ \fbi$ at $3\sigma$
statistical significance for 110 GeV $<m_h<140$ GeV in the Standard Model.
This corresponds to a $h\mu\mu$ coupling determination at about $15\%$ accuracy
assuming $ht\bar t,\ hb\bar b$ couplings SM-like.  The observation becomes
more promising in the MSSM for $\tan\beta>8$ and $M_A<130$.  
\end{abstract}

\maketitle

\vskip 0.5cm
\noindent
{\it I. Introduction}
 
The Higgs mechanism is widely believed to be responsible for the electroweak
gauge symmetry breaking and possibly for the fermion mass generation. Searching
for Higgs bosons have thus become high priority in future collider experiments,
and the Large Hadron Collider (LHC) that is under construction at CERN has the
promise to discover the Standard Model (SM) Higgs boson or the counterparts in
theories with a Supersymmetric extension \cite{LHC}. 
After the initial discovery,
it would be more important and more challenging to understand the properties of
the Higgs bosons, in particular their couplings to the SM particles. Precisely
because of the role of the Higgs bosons in the mass generation mechanism, they
couple to the SM particles proportional to their masses. 
Such characteristics of
Higgs bosons should be thoroughly tested at collider experiments.

At the LHC, the SM Higgs ($h$) couplings to 
$WW,\ ZZ,\ t\bar t,\ \tau^-\tau^+$,
as well as the loop-induced couplings to $gg,\ \gamma\gamma$ can all
be directly probed by combining several production and decay channels
\cite{LHC,dieter}. At an $e^+e^-$ linear collider, this list can be extended to
include $b\bar b,\ c\bar c$ \cite{nlc}.  The next channel anticipated would be
$\mu^+\mu^-$, which would be the first observation for a Higgs boson
to decay to second-generation fermions. 
It is in fact very important to explore this channel. 
First, it is necessary to confirm the proportionality of $m_\mu$ 
for the Higgs coupling $h\mu\mu$ as predicted by the SM.  
Secondly, this channel may be sensitive to new physics
such as non-universality between $I_3=-{1\over 2}$ fermions 
in certain classes of SUSY models or induced by 
radiative corrections. Last but not least, the concept of a muon collider
Higgs factory relies on the $h\mu\mu$ coupling \cite{mc}.
Due to the rather small branching fraction for $h\to \mumu$, it
will be very challenging to observe this channel in collider experiments. 
An early study of this channel at the LHC based on the weak-boson fusion, 
$WW,ZZ\to h\to \mumu$, was carried out \cite{til}. 
There would be about $1-2\sigma$ statistical effect 
obtained at the LHC for
an integrated luminosity of $300\ \fbi$. At a future $e^+e^-$
linear collider with $\sqrt s\ge 800$ GeV and an integrated luminosity 
of 1000 $\fbi$, it may be possible to
reach a $15\%$ measurement for the $h\mu\mu$ coupling \cite{eehmumu}.

We note that the leading Higgs production at the LHC is via the gluon
fusion, yielding the process
\begin{equation}
gg\to h\to \mumu.
\label{ggh}
\end{equation}
Because of the large Higgs production rate at the LHC and the very clean
experimental signature of $\mumu$, we are thus motivated to explore 
this channel and wish to improve the observability of $h\to \mumu$.
It is important to note that the cross section for the process of
Eq.~(\ref{ggh}) is proportional to 
$\Gamma(h\to gg)\times BR(h\to \mumu)$. The partial width 
$\Gamma(h\to gg)$ is dominated by the top-quark loop in the SM,
and will receive contributions from new particles beyond the SM
that are
colored and couple to $h$ significantly. The branching fraction
$BR(h\to \mumu)$ may also deviate from the SM prediction, especially
if the new physics contribution breaks the universality
between the muon and the $b$-quark. We thus expect that the process of
Eq.~(\ref{ggh}) would be sensitive to new physics beyond the SM.
As a concrete example, we also include discussions in the minimal 
supersymmetric Standard Model (MSSM) in our analysis. 

\vskip 0.5cm
\noindent
{\it II. Signal and Background Studies} 

We study the signal process of Eq.~(\ref{ggh}) at the LHC
with the center-of-mass energy $\sqrt{s}=14$ TeV. 
We first note that in the SM, the rate of 
$h\to \mumu$ channel becomes vanishingly small
once the $h\to WW^*,\ ZZ^*$ channels are open. We thus concentrate 
on the mass range
\begin{equation} 
    110\ {\gev} < m_h < 140\ \gev.  
\label{mhrange}
\end{equation}
This is also the mass window favored by supersymmetric extensions of the SM.

Because of the very clean final state of $\mumu$, we will consider 
the inclusive
channel.  We use CTEQ4M structure functions \cite{cteq4m}.  We calculate the
signal cross section by normalizing the rate with respect to the output of the
packages HIGLU \cite{higlu} for the SM and HDECAY \cite{hdec} for MSSM. The
NLO $K$-factor for the Higgs production via gluon fusion is about 2.5 at the LHC
energy, larger than that previously reported (around 1.5) \cite{higlu} due to
increased accuracy in gluon parton density functions at low $x$.  The Higgs
production is treated in the narrow width approximation, which is fully
justifiable because the physical Higgs width is much less than the experimental
detector resolution.  The irreducible SM background comes from the Drell-Yan
production
\begin{equation} q\bar q \to Z^*,\gamma^*\to \mumu.  \end{equation}
We have normalized the background cross section with respect to that with QCD
corrections \cite{dy}. We stress that we will know this DY background 
quite well from the direct measurement at the LHC experiments.

We simulate the experimental detector coverage by imposing the kinematical 
cuts on both muons
\begin{equation} 
p_T > 20\ \gev, \quad \eta < 2.5.  
\label{basic}
\end{equation}
The detector smears the muon momentum approximately to a Gaussian form 
of a width $\sigma=1.6\ \gev$ \cite{LHC,til}.  To optimize the statistical
significance, we find that the maximum $S/\sqrt{B}$ occurs when this invariant
mass window is $\pm 1.4\sigma$ around the peak. We thus take the invariant 
mass as
\begin{equation}
m_h-2.24\ {\gev} < m(\mumu) < m_h+2.24\ {\gev},
\label{mass}
\end{equation}
which captures 84\% of the signal.  An identification efficiency of $90\%$ for
each muon is also included in our analysis.

\begin{table}[tbh]
    \setlength{\tabcolsep}{5mm}
    \begin{tabular}[t]{c|c|c|c|c}
        &
        \multicolumn{2}{c|}{Gluon fusion} & \multicolumn{2}{c}{$W$ boson fusion} \\
        \cline{2-5}
        \raisebox{1.5ex}[0pt]{$m_h (\gev)$} & signal & background & signal & 
            \multicolumn{1}{c}{background} \\
        \hline
        115 & 4.50 & 2085 & 0.092 & 0.82 \\
        120 & 3.89 & 1441 & 0.081 & 0.62 \\
        130 & 2.63 & 821  & 0.062 & 0.40 \\
        140 & 1.51 & 526  & 0.037 & 0.28 \\
    \end{tabular}
    \caption{
        SM cross sections in fb for both gluon fusion and 
weak-boson fusion signals, and the corresponding backgrounds after all cuts.
The cuts used are in Eqs.~(\ref{basic}) and (\ref{mass}).  
A $90\%$ muon identification efficiency factor is included.  
        The weak-boson fusion results are taken from \cite{til}.
    } 
\end{table}

We first give the signal and background cross sections in 
Table I after the cuts and the efficiency factor as discussed
above for the Higgs mass  range of interest in Eq.~(\ref{mhrange}). 
For comparison, results for weak-boson fusion are
also listed, as taken from Ref.~\cite{til}. Although the signal
rate is larger for gluon fusion than that for weak-boson fusion
by more than a factor of 40, 
the background here is substantially larger as well.
However, we emphasize that the Drell-Yan background 
will be precisely measured at the LHC experiments. 
The systematic effects due to theoretical uncertainties
will be minimal.
The signal we
are looking for is a $\mumu$ mass peak at an approximately known location
on a very well-measured, nearly flat background.
In contrast, the weak-boson fusion process yields a 
signal-to-background ratio of better than $10\%$. The further
challenge is to understand systematic errors better.

Table II summarizes our SM results 
combining both ATLAS and CMS detectors. 
We first give the delivered luminosity needed
to reach a 3$\sigma$ observation of the signal, which corresponds
to the cross section determination to about $33\%$ accuracy,
as estimated by $\sqrt{S+B}/S$.
If we assume that the couplings of $ht\bar t$ and $hb\bar b$
are known to be SM-like, then the above accuracy of $h\to \mumu$ 
branching fraction determination translates to
the $h\mu\mu$ coupling determination to about $17\%$. 
We see that combining both signal channels and with two detectors,
the typical luminosity needed is about 250 $\fbi$ to reach
this level of accuracy. 
With 300 fb$^{-1}$ delivered to each detector, 
one can reach a 3.5$\sigma$ observation statistically as shown
in the last three columns
in Table II. This corresponds to $h\to \mumu$ branching 
fraction determination to about $29\%$ accuracy, 
or the $h\mu\mu$ coupling determination to 14$\%$,
assuming $ht\bar t,\ hb\bar b$ couplings SM-like.
With extended running or luminosity upgrades, a $5\sigma$
observation may be a reasonable expectation.
Overall, we see that our results for gluon fusion is
quite comparable to the earlier study from weak-boson
fusion, and that combining these two channels can significantly
improve the observability of $h\to\mumu$ at the LHC experiments.

\begin{table}[tbh]
    \begin{tabular}[t]{c|c|c|c|c|c|c}
            &
        \multicolumn{3}{c|}{Luminosity required for 3$\sigma$ observation ($\fbi$)} &
        \multicolumn{3}{c}{Significance for $300\ \fbi$} \\
        \cline{2-7}
        \raisebox{1.5ex}[0pt]{$m_h (\gev)$} & $W,g$ Combined & $g$ fusion & 
$W$ fusion & $W,g$ Combined 
            & $g$ fusion & $W$ fusion \\
        \hline
        115 & 238 & 464 & 489 & 3.37 & 2.41 & 2.35 \\
        120 & 227 & 430 & 482 & 3.45 & 2.51 & 2.37 \\
        130 & 267 & 535 & 532 & 3.18 & 2.25 & 2.25 \\
        140 & 531 & 1047& 1076& 2.26 & 1.61 & 1.58 \\
    \end{tabular}

    \caption{
        The SM results for $h\to \mumu$ signal from gluon fusion and 
weak-boson fusion and the DY background, 
        combining the ATLAS and CMS detectors. 
The cuts used are in Eqs.~(\ref{basic}) and (\ref{mass}).  
A $90\%$ muon identification efficiency factor is included.  
        The weak-boson fusion results are taken from \cite{til}.
    }
\end{table}

\begin{figure}[tbh]
    \includegraphics[scale=0.45]{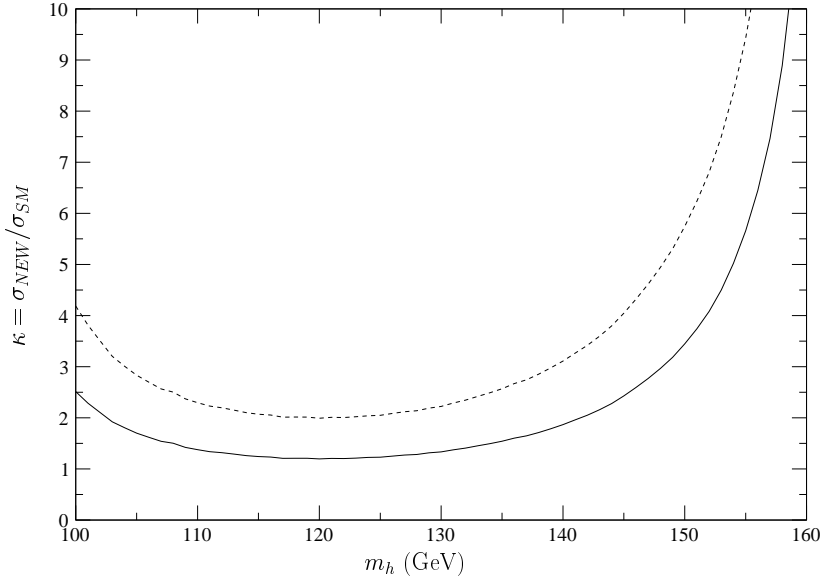}
    \caption{The enhancement factor $\kappa$ over the SM rate
required to observe the 
$gg\to h\to \mumu$ signal at the $3\sigma$ (solid) and  
$5\sigma$ (dashed) level with $300\ \fbi$ delivered
    luminosity, including both the ATLAS and CMS detectors.}
\label{kap}
\end{figure}

Many theories beyond the SM can lead to significant enhancement
for the channel $gg\to h\to \mumu$ 
and thus the signal observation may be easier.
As a model-independent generic argument, we study the cross section 
enhancement factor ($\kappa$) over the gluon fusion
channel in the SM. The curves in Fig.~\ref{kap} show the enhancement
factor $\kappa$ needed to reach a $3\sigma$ (solid) and 
$5\sigma$ (dashed) 
signal via the gluon fusion channel alone versus the Higgs mass $m_h$, 
with both detectors and for a delivered
luminosity of $300\ \fbi$. We note that for a low Higgs mass
$m_h < 110$ GeV, the signal observation is difficult primarily
because of the overwhelmingly large DY background from the tail
of the $Z$-pole. On the other hand, for $m_h>140$ GeV, the signal
observation becomes increasingly difficult due to the fact that
the $h\to \mumu$ channel dies away after the opening of $h\to W^*W,
Z^*Z$ channels. 
For the mass range of our current interest
$110\ {\gev} <m_h< 140\ \gev$, an enhancement factor of
$\kappa\sim 1.2-2$ is needed for a $3\sigma$ signal observation
and $\kappa\sim 2.1-3.3$ for a $5\sigma$ signal observation, 
for a delivered luminosity of $300\ \fbi$.
To present this in another way, given the $\kappa$ factor, the 
luminosity required to observe this channel at
an $S$ significance level is simply given by:
\begin{equation}
    \mathcal{L} = S^2\frac{\sigma_B}{\kappa^2 \sigma_S^2}
\end{equation}
where $\sigma_B$ and $\sigma_S$ are the
background and SM signal cross sections, respectively,
presented in Table I.
As a concrete example, we will study the enhancement factor
in MSSM next.

\vskip 0.5cm
\noindent
{\it III. Minimal Supersymmetric Standard Model} 

In MSSM there are two Higgs doublets, resulting in 5 physical Higgs
states. The relevant parameters are $\tan\beta$, the ratio of
the two vacuum expectation values, and $M_A$, the mass of
the CP-odd Higgs state. The $\mumu$ mode via gluon fusion may be 
significantly enhanced in MSSM. 
First of all, there are SUSY particles such as stops and
sbottoms to contribute in the loop. However, there are also
subtle cancellations among the diagrams \cite{abd}. Secondly,
for large $\tan\beta$, the $b$ quark and sbottom contributions 
can be significant. Thirdly , there may be direct
contribution from $A,H\to \mumu$ \cite{LHC,vb}.

We consider the maximal stop quark mixing scenario \cite{maxstop},
defined by the stop mixing parameter 
$X_t \equiv A_t - \mu \cot\beta = \sqrt{6} M_{SUSY}$,
where $A_t$ is the soft SUSY breaking top Yukawa coupling, 
$\mu$ is the dimensionful
Higgs mixing parameter, and $M_{SUSY}$ is the mass of the squarks (where all
squarks are assumed to be degenerate in mass).
The maximal stop mixing scenario gives us larger $m_h$ to be
consistent with the current LEP2 Higgs mass bound \cite{mabound}.
This also happens to lead to a large production cross section
of $gg\to h$ in the low $M_A$ and large $\tan\beta$ limit. 
For our simulations in the MSSM we have chosen the parameters
$M_{SUSY}=1$ TeV, $\mu=300\ \gev$, and $A_U=A_L=A_D=1.5$ TeV.  The $h\mu\mu$
coupling is insensitive to these parameters.

\begin{figure}[tbh]
    \includegraphics[scale=0.45]{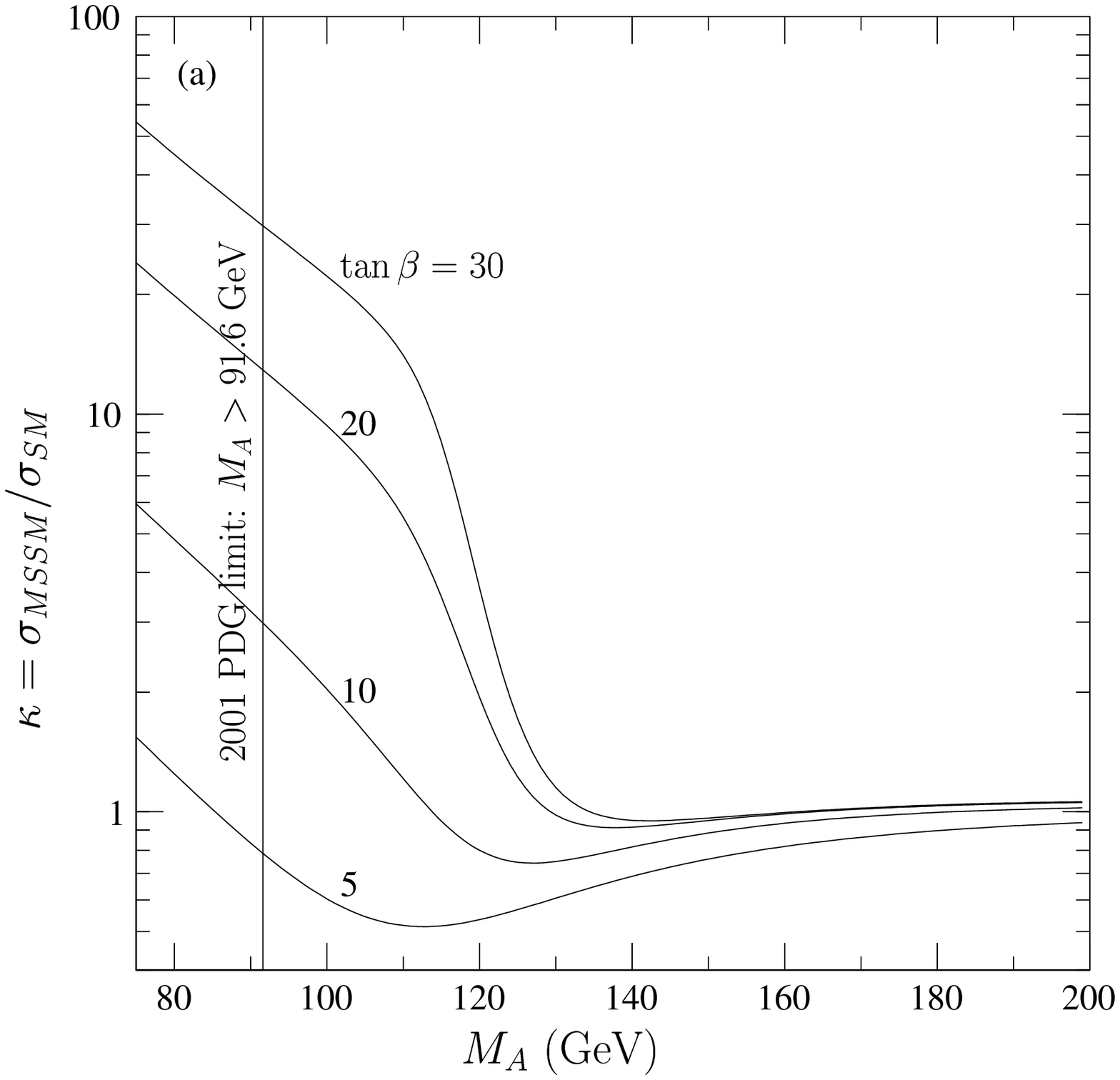}
    \includegraphics[scale=0.45]{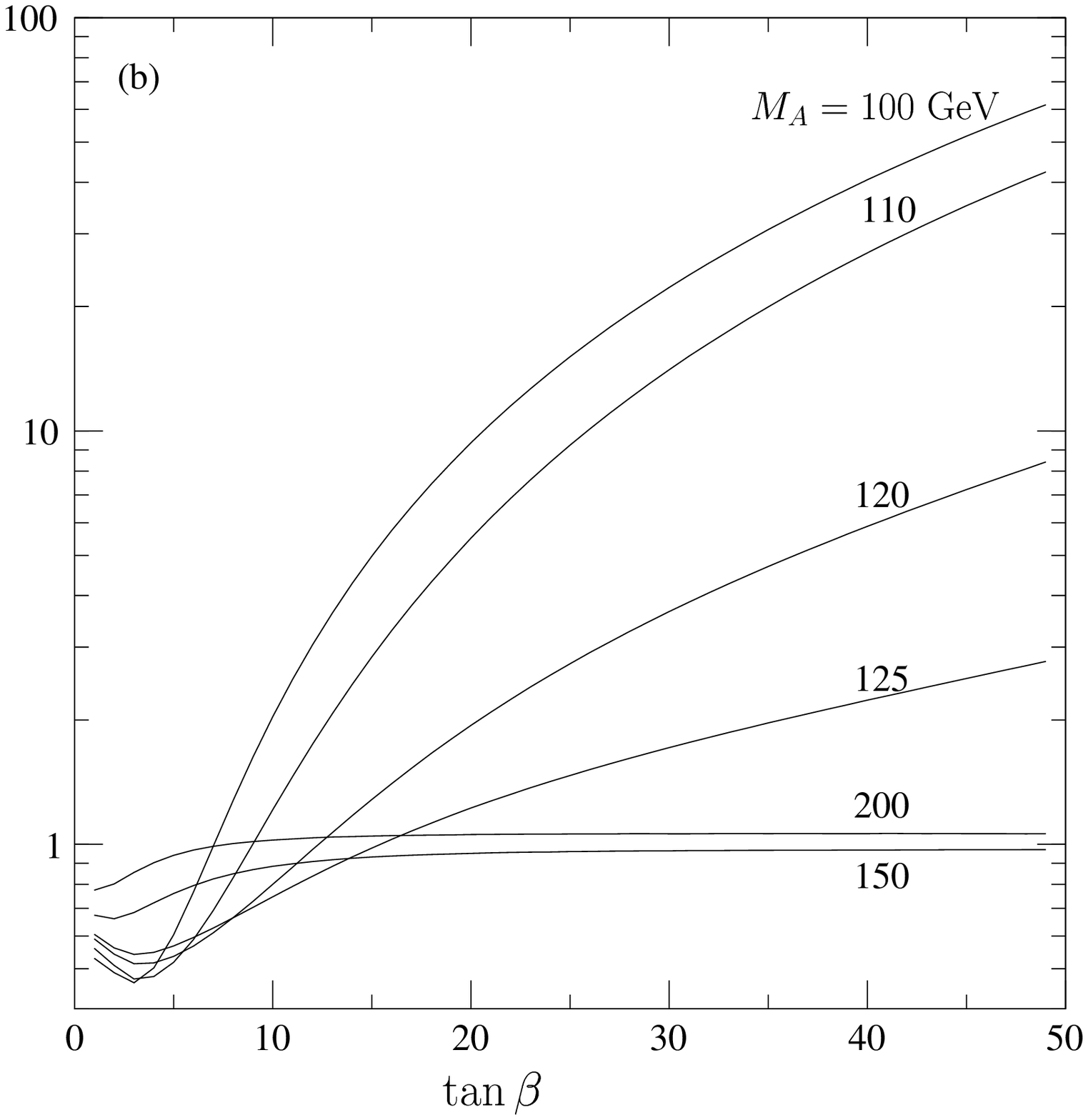}
    \caption{The enhancement of $h$ production of 
the MSSM relative to the SM in the maximal stop
    mixing scenario as a function of (a) $M_A$ (left) and (b)
    $\tan\beta$ (right). The curves are labeled by
    their value for $\tan\beta$ in (a) and $M_A\ (\gev)$ in (b).} 
\label{susy}
\end{figure}

In Fig.~\ref{susy}, we present the enhancement factor 
$\kappa$ for $gg\to h\to \mumu$ in MSSM. 
We see that for low $M_A$ and large $\tan\beta$
the enhancement can be substantial, as large as a factor of
20$-30$ at the edge of the $M_A$ exclusion of 91 GeV \cite{mabound}.
This does not include the possible contribution from $H,A$ decay yet. 
As anticipated, in the heavy limit $M_A>2M_Z$, 
we recover the Standard Model result ($\kappa=1$).
Based on comparison to the SM case as discussed for Fig.~\ref{kap}, 
we conclude that
the $h\to \mumu$ channel in MSSM can be observed at the $3\sigma$ level
or better for $M_A<130$ GeV and for $\tan\beta>8$ with $300\ \fbi$ 
luminosity delivered.

\vskip 0.5cm
\noindent
{\it IV. Conclusions} 

We have studied the Higgs decay to $\mumu$ via gluon fusion process at the LHC
in the SM and MSSM. We found that this channel is quite comparable and
complementary to the channel from the weak-boson fusion.
By including both the gluon fusion channel
and the weak-boson fusion channel, and by including the ATLAS and
CMS detector, the LHC with 300 $\fbi$ can observe the $h\to \mumu$ 
to a statistical significance of $3\sigma$ over a Higgs mass range of 
110 GeV$<m_h<$140 GeV. This corresponds to the $h\mu\mu$ coupling
determination about $14\%-17\%$ accuracy if assuming $ht\bar t,\ hb\bar b$
couplings SM-like.
If nature has chosen large $\tan{\beta}$ as is preferred in a large class
of SUSY models and as favored by present experimental limits,
and in addition $M_A<130$ GeV, 
we might easily observe this channel at the LHC and determine
the branching fraction to an accurate level.

{\it Acknowledgments}:
We thank Uli Baur, Tilman Plehn, Dave Rainwater and Dieter Zeppenfeld for
discussions.  This work was supported in part by a DOE grant No.
DE-FG02-95ER40896 and in part by the Wisconsin Alumni Research Foundation.

\end{document}